\documentclass[aps,prb,showpacs,superscriptaddress,twocolumn]{revtex4-1}
\usepackage{graphics} 
\usepackage{amsmath}
\usepackage{amssymb}
\usepackage{dcolumn}
\usepackage{bm}
\usepackage{graphicx}  
\usepackage[utf8]{inputenc}
\usepackage{url}
\usepackage{wrapfig}
\usepackage[caption=false]{subfig}
\usepackage{xcolor}
\usepackage[T1]{fontenc}

\begin{document}

\newcommand*{\citen}[1]{%
  \begingroup
    \romannumeral-`\x 
    \setcitestyle{numbers}%
    \cite{#1}%
  \endgroup   
}

\title{Magnetic anisotropy in Fe/U and Ni/U bilayers}
\author{E.R. Gilroy}\altaffiliation{Current address: Department of Materials Science and Engineering, Sir Robert Hadfield Building, Mappin Street, Sheffield, S1 3JD, United Kingdom}
\affiliation{H. H. Wills Physics Laboratory, University of Bristol, Tyndall Avenue, Bristol BS8 1TL, United Kingdom}

\author{M.-H. Wu}
\affiliation{H. H. Wills Physics Laboratory, University of Bristol, Tyndall Avenue, Bristol BS8 1TL, United Kingdom}
\author{M. Gradhand}
\affiliation{H. H. Wills Physics Laboratory, University of Bristol, Tyndall Avenue, Bristol BS8 1TL, United Kingdom}
\affiliation{Institut f\"ur Physik, Johannes‐Gutenberg‐Universit\"at Mainz, Staudingerweg 7, 55128 Mainz, Germany}
\author{R. Springell}
\affiliation{H. H. Wills Physics Laboratory, University of Bristol, Tyndall Avenue, Bristol BS8 1TL, United Kingdom}
\author{C. Bell}
%
\affiliation{H. H. Wills Physics Laboratory, University of Bristol, Tyndall Avenue, Bristol BS8 1TL, United Kingdom}
\begin{abstract}
Magnetometry measurements of Fe/U and Ni/U bilayer systems reveal a non-monotonic dependence of the magnetic anisotropy for U thicknesses in the range 0$\,$nm - 8$\,$nm, with the Fe/U bilayers showing a more prominent effect as compared to Ni/U. The stronger response for Fe/U is ascribed to the stronger 3$d$-5$f$ hybridization of Fe and U. This non-monotonic behaviour is thought to arise from quantum well states in the uranium overlayers. Estimating an oscillation period from the non-monotonic data, and comparing it to Density Functional Theory calculations, we find that wave vector matches to the experimental data can be made to regions of high spectral density in (010) and (100) cuts of the electronic structure of $\alpha$-U, consistent with the measured texture in the films. Unexpectedly, there are also indications of perpendicular magnetic anisotropy in a subset of Fe/U samples at relatively large U thickness. 

\end{abstract}

\maketitle

\section{Introduction}
Spin-orbit coupling (SOC) profoundly affects the band structure of a material, leading to many exotic phenomena such as topological insulating states\cite{qi2011topological}, and Rashba spin splitting\cite{manchon2015new}. In magnetic materials and spintronic systems, large SOC is at the heart of magnetic anisotropy, the spin Hall effect (SHE)\cite{hirsch1999spin,valenzuela2006direct,saitoh2006conversion,liu2012spin, hoffmann2013spin}, and the Dzyaloshinskii-Moriya interaction observed in magnetic-heavy metal structures\cite{heinze2011spontaneous}. In the simplest picture the SOC of a material increases $\propto$ Z$^4$, where Z is the atomic number\cite{tserkovnyak2005nonlocal}. Therefore there has recently been intense focus on spintronic systems containing relatively heavy non-magnetic metals such as Pt, Au and Ir, in which variety of effects can be observed. For example, by growing thin films of these heavy metals next to magnetic materials, the spin currents produced by the SHE when a charge current is passed in the heavy metal (HM) layer can be used for spin transfer torque switching of ferromagnetic layers at relatively low current densities\cite{liu2012spin}. At the same time, however, heavy metals cause enhanced spin damping in the ferromagnetic layer\cite{Mizukami2002,zhang2017influence}, and are susceptible to proximity-induced magnetism. Induced moments have been detected in systems such as Fe/{Pt}\cite{antel1999induced} and Co/Pt\cite{Huang2012} and are thought to inhibit the efficiency of spin current detection through the inverse SHE\cite{Huang2012}. Understanding the influence of the interfacial induced moment and the large SOC is an important challenge. 

The presence of an overlayer on a ferromagnetic (FM) film can have significant influence on the magnetic anisotropy of the system. It is well documented that in thin FM/HM structures there can be an emergence of perpendicular magnetic anisotropy, as the magnetization of the ferromagnet is pulled out-of-plane by an interfacial anisotropy contribution \cite{miura2013first, Hashimoto1989}. Additionally, when the thickness of the overlayer is altered, oscillations in the magnetic anisotropy can be detected \cite{ortega1992quantum,manna2013effect}. These oscillations in anisotropy are thought to originate from quantum well states,which arise due to confinement of electrons at the interface\cite{bruno1995theory,himpsel1998electronicIBM,cinal2001analysis,przybylski}. 

In the context of these fascinating effects, the study of uranium - with the largest Z for a naturally occurring element - is of considerable interest as a HM in FM/HM heterostructures. Previous X-ray magnetic circular dichroism (XMCD) measurements at the U $M_{4,5}$ edges have observed negligible induced moment in U when grown on Ni and Co, but a relatively large moment when grown on Fe\cite{FeNiCo,xmcd}. Hence by varying the FM layer in FM/U heterostructures it may be possible to disentangle the role of the induced moment and that of the large SOC in the U. This paper studies Fe/U and Ni/U in bilayer systems, i$.$e$.$ with and without an induced moment in the U, respectively, focusing on the effect of the U overlayer thickness on the magnetic anisotropy of the FM film.

\section{Methods and structural characterization}
\begin{figure}[h!]
\includegraphics[scale = 0.4]{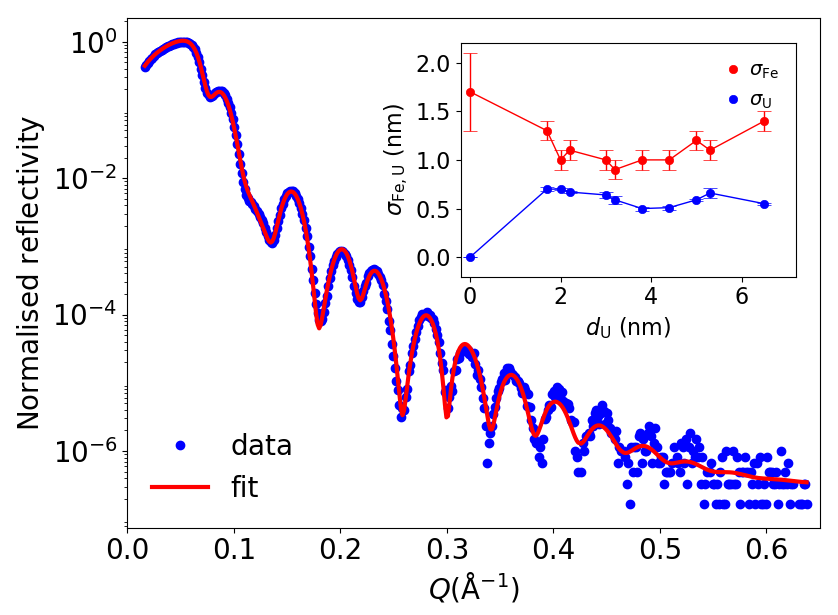}
        \caption{[Color online] Example normalised low angle x-ray reflectivity data (symbols) for the Fe/U samples. Solid line is a best fit using the GenX software, in this case giving $d_{\mathrm Fe} = 8.3 \pm 0.1\,$nm and $d_{\mathrm{U}} = 6.5 \pm 0.1\,$nm. Root-mean-square roughness of Fe and U, ($\sigma_{\mathrm{Fe}}$ and $\sigma_{\mathrm{U}}$ respectively), determined from the fit against $d_{\mathrm U}$ are shown in the inset. Similar data were obtained for the Ni series, and are discussed in Appendix \ref{appenda}.}   \label{fig:xrr}
\end{figure}

The samples were grown by d$.$c$.$ magnetron sputtering at room temperature in an ultra-high vacuum chamber with a base pressure of $<2\times 10^{-9}\,$mbar. The argon sputtering pressure was held constant at ($7.0\pm 0.1) \times 10^{-3}\,$mbar for all layers. The substrates were 10$\,$mm $\times$ 10$\,$mm $\times$ 0.7$\,$mm on Corning glass polished to optical grade. The sample structure was glass/FM/U/Nb, where the FM layer was Fe or Ni. The Fe thickness fixed at $\sim$8.5$\,$nm, the Ni at $\sim$11$\,$nm. To avoid oxidation a cap of Nb with thickness fixed at $\sim$8$\,$nm was used for the Fe series, and  $\sim$10$\,$nm thick for the Ni series. The uranium thickness, $d_{\mathrm U}$, was varied in the range $0-8\,$nm. The deposition rates were 0.017, 0.03, 0.085 and 0.064$\,$nm/s for the Fe, Ni, U, and Nb respectively. Deposition rates were calibrated and samples were characterised using X-ray reflectivity (XRR). The GenX reflectivity software\cite{Bjorck:aj5091,GenX} was used to determine the thicknesses and roughness of each sample. The errors on thickness produced by GenX are given as the value required to change the best fit figure of merit (FOM) by $\pm 5\%$. The FOM used here gave equal weighting to both high and low intensity points. Example XRR data are shown in Fig$.$ \ref{fig:xrr}, showing excellent agreement between the model and the fit. Over the range of U thicknesses both the Fe and Ni the roughness did not significantly, other than at $d_{\mathrm U} = 0\,$nm where the interface is FM/Nb (see inset of Fig$.$ \ref{fig:xrr}). The average roughness values $\sigma_{\mathrm{Fe}} = 1.1\,$nm and $\sigma_{\mathrm{Ni}} = 1.3\,$nm are typical for room temperature sputtered metal thin films.

\begin{figure}[h!]
\includegraphics[scale = 0.4]{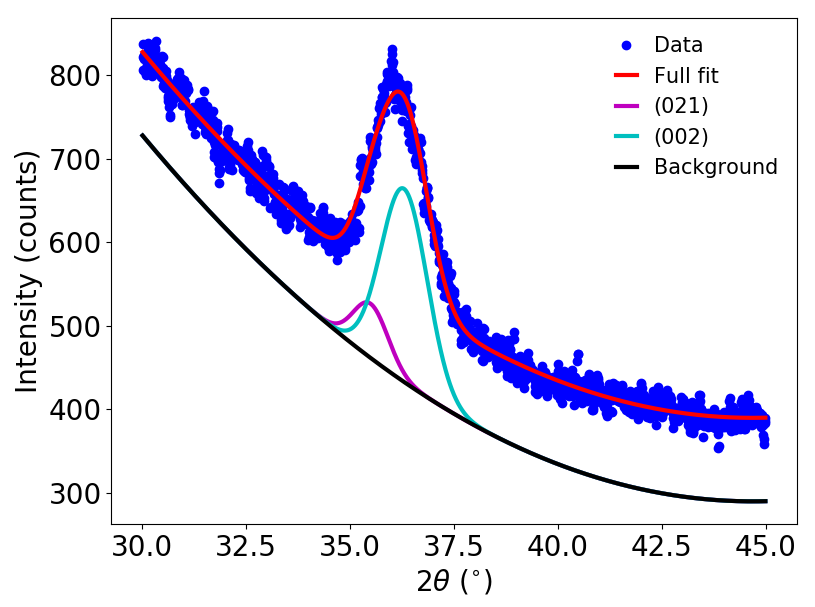}
        \caption{[Color online] High angle x-ray diffraction data for the Fe series with $d_{\mathrm{U}} =3.8\,$nm. The solid red line is the total fit of three Gaussians fixed at the uranium triplet positions. Magneta and cyan lines are fits to (021) and (002) reflections respectively. The black line is a second order polynomial fit to the background. } 
        \label{fig:xrd}
\end{figure}
X-ray diffraction data were also taken, both in $\theta-2\theta$ and grazing incidence geometries, to examine the possibility of texture in the polycrystalline samples. For Fe/U there was evidence of the U layer being oriented with predominantly the [001] direction normal to the plane, with a smaller fraction of the layer also oriented in the [011] (see Fig$.$ \ref{fig:xrd}), and Ni/U only exhibiting [001] texture. This suggests that in both cases the U overlayer is textured.

The samples were divided to 5$\,$mm $\times$ 5$\,$mm pieces with a diamond saw for room temperature vibrating sample magnetometry (VSM). Magnetic moment $M$ $vs$ applied field $H$, hysteresis loop measurements were carried out with $H$ applied both in the plane and perpendicular to the plane of the samples. The in-plane angle $\theta$ ranged from $-10^{\circ}$ to 190$^{\circ}$ in 10$^{\circ}$ steps, relative to an arbitrary in-plane axis defined parallel to the main axis of the sputtering chamber.

Density Functional Theory calculations were carried out using a fully relativistic Korringa-Kohn-Rostoker Greens function method \cite{gradhand_prb2009} extended to include calculations for the Bloch spectral function \cite{saunderson_prb2020} Further details of the calculations for the uranium crystal can be found in Ref$.$ \citen{wu_prb2020}.

\section{Results and discussion}
\subsection{In-plane magnetometry}
\begin{figure}[h]
\includegraphics[scale = 0.4]{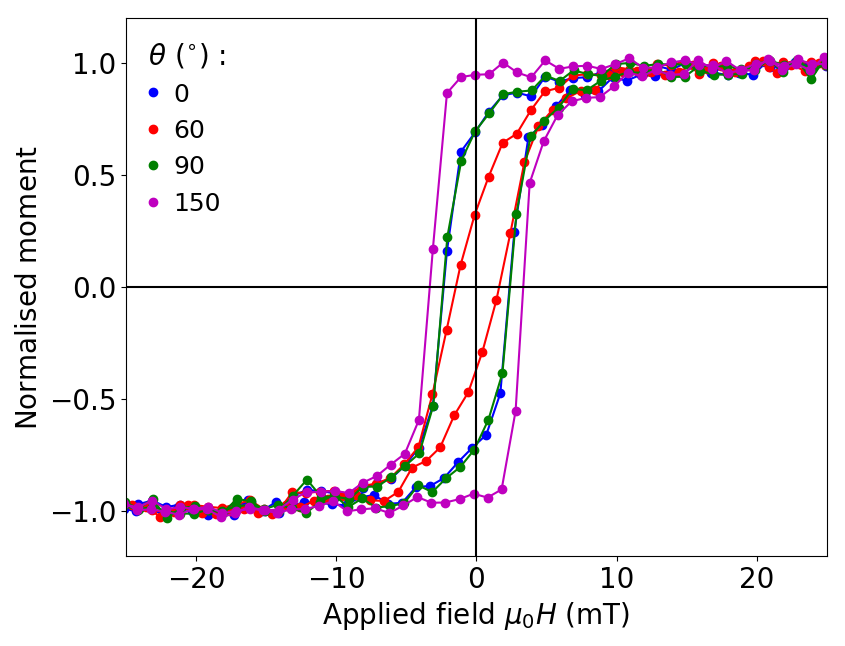}
        \caption{[Color online] Room temperature in-plane $M(H)$ loops for sample with FM = Fe and $d_{\mathrm{U}} = 6.5\,$nm. Moment $M$ is normalised to the saturation value. Relatively hard (easy) axis behavior is observed for $\theta = 60^{\circ} (150^{\circ})$. A subset of the $M(H,\theta)$ data are shown for clarity. Lines are guides to the eye.}
        \label{fig:angularloops}
\end{figure}

\begin{figure}[h]
\includegraphics[scale = 0.4]{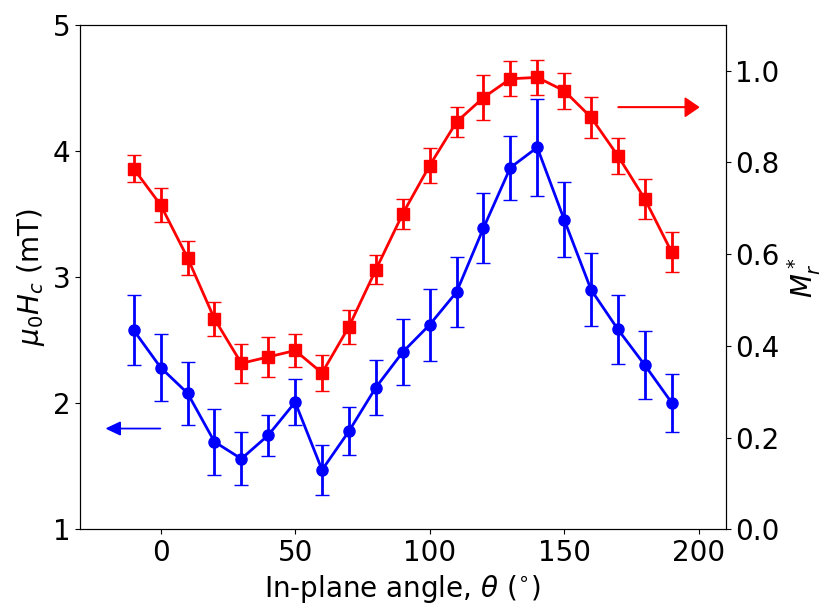}
        \caption{[Color online] In-plane angular dependence of the coercive field $\mu_0 H$, and normalised remnant moment $M_r^*$ for FM = Fe and $d_{\mathrm{U}} = 6.5\,$nm. Lines are a guide to the eye.}
        \label{fig:angularvsm}
\end{figure}

Typical in-plane $M(H,\theta)$ data are shown in Fig$.$ \ref{fig:angularloops} for the case of FM = Fe and $d_{\mathrm{U}} = 6.5\,$nm. There are clear changes in anisotropy with in-plane angle $\theta$. Figure \ref{fig:angularvsm} shows the full evolution of the anisotropy through both the coercive field, $H_c$, and normalised remnant moment, $M_r^*$ with angle. The data show clear uniaxial anisotropy, with the easy and hard axes situated at 140$^\circ$ and 60$^\circ$ respectively. Small peaks in $H_c$ and $M_r^*$ seen around 50$^{\circ}$ are likely due to stray fields within the sputtering chamber arising from the magnets within the sputtering guns. It is most likely that the uniaxial anisotropy arises from the off-normal angle of incidence of the sputtered atoms relative to the substrate\cite{ono1993texture,chowdhury2014controlling}.

The range of the coercive field, $H_c$, as a function of angle, as well as the average $H_c$ of each sample is used as a way of quantifying the anisotropy of the  system. The range is defined as 
$ H_c^{\mathrm{Range}} = H_{c(\mathrm{max})} - H_{c(\mathrm{min})}, $ where $H_{c(\mathrm{max})}$ and $H_{c(\mathrm{min})}$ are the maximum and minimum values of $H_c(\theta)$, respectively. The average $H_c$ is given by $H_c^{\mathrm{Ave}} =\frac{1}{n}\displaystyle\sum^{n}_{i=1}H_{c}^{i},$ where $n$ is the total number of angular scans.


\begin{figure}[h]
\includegraphics[scale = 0.4]{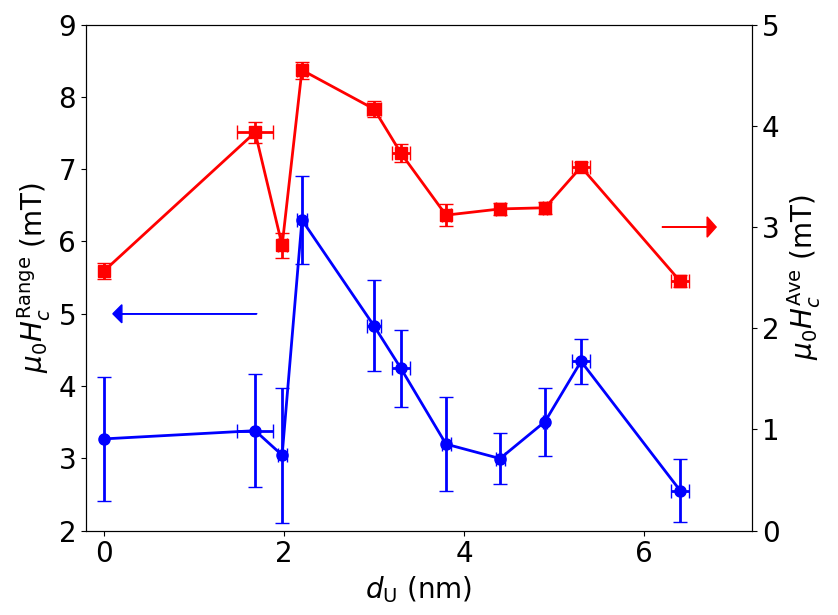}
        \caption{Range of coercive field $\mu_0 H_c$ (left axis) and average $H_c$ (right axis) vs $d_{\mathrm{U}}$ for Fe series. Clear non-monotonic behavior observed for both parameters. Lines are a guide to the eye.} 
        \label{fig:hcrange_fe}
\end{figure}

Figure \ref{fig:hcrange_fe} (left axis) illustrates the development of $H_c^{\mathrm{range}}$ with $d_{\mathrm{U}}$ for the Fe series. As  $d_{\mathrm U}$ increases, the range displays clear non-monotonic behaviour. At  $d_{\mathrm U}$ $\sim$ 2$\,$nm, $ H_c^{\mathrm{Range}}$ is enhanced by a factor of two in comparison to  $d_{\mathrm U}$ = 0$\,$nm. The right hand axis in Fig$.$ \ref{fig:hcrange_fe} shows $H_c^{\mathrm{ave}}$ for each sample. This closely follows the same non-monotonic form as $H_c^{\mathrm{range}}(d_{\mathrm{U}})$.

\begin{figure}[h]
\includegraphics[scale = 0.4]{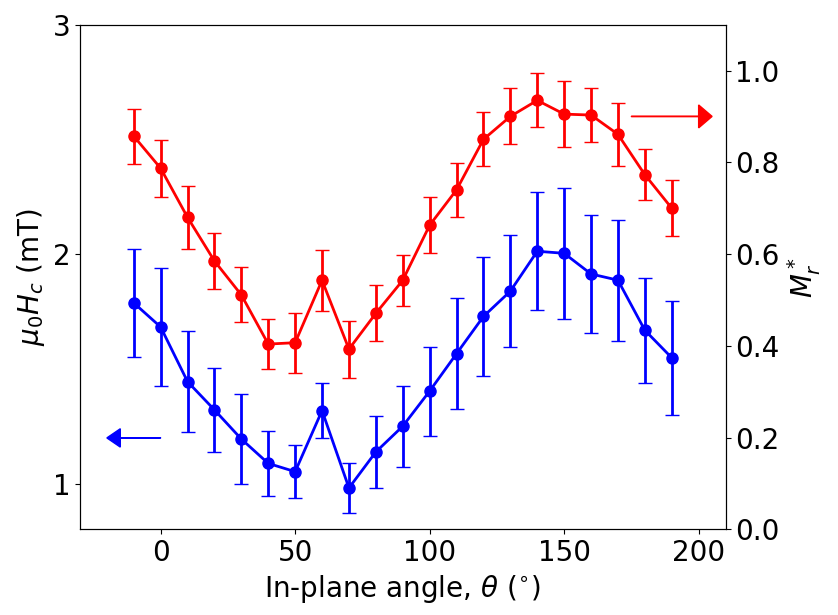}
        \caption{[Color online] In-plane angular dependence of the coercive field $\mu_0 H$, and  normalised remnant moment $M_r^*$ for a Ni bilayer with $d_{\mathrm{U}} = 0.6\,$nm.  Lines are a guide to the eye.}
        \label{fig:angularvsm_ni}
\end{figure}

\begin{figure}[h]
\includegraphics[scale = 0.4]{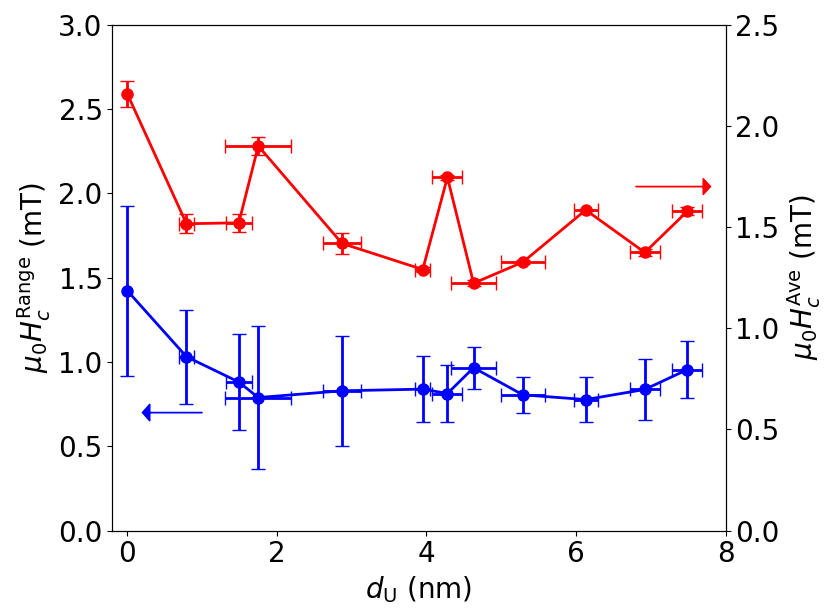}
        \caption{Range of coercive field $\mu_0 H_c$ (left axis) and average $H_c$ (right axis) vs $d_{\mathrm{U}}$ for Ni series. Non-monotonic behavior observed for 
        average coercive field (right axis). Lines are a guide to the eye.} 
        \label{fig:hcrange_ni}
\end{figure}

Next we examine the FM = Ni samples as a comparison with the Fe samples. The samples showed very similar uniaxial anisotropy to the Fe samples, strongly suggesting that the uniaxial anisotropy is induced through the sputtering process. Both the $H_c^{\mathrm{range}}$ and $H_c^{\mathrm{ave}}$ were about a factor of four and two smaller than the Fe counterpart respectively, as shown in Figs$.$ \ref{fig:angularvsm_ni} and \ref{fig:hcrange_ni}. The Ni series $H_c^{\mathrm{ave}}(d_{\mathrm{U}})$ showed a clear non-monotonic dependence not dissimilar to the Fe series, although for Ni the $H_c^{\mathrm{range}}(d_{\mathrm{U}})$ does not clearly mirror the $H_c^{\mathrm{ave}}(d_{\mathrm{U}})$ data in the way found for the Fe series data in Fig$.$ \ref{fig:hcrange_fe}. The reason for this disparity is likely related to the low switching field of Ni and the small variation of $H_c$ with angle.


\begin{figure}[h]
 \includegraphics[scale = 0.4]{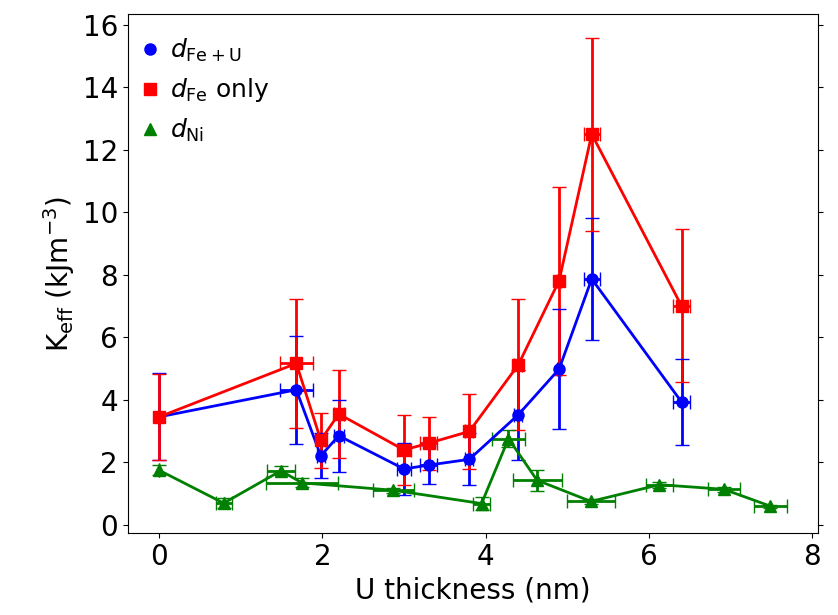}
        \caption{[Color online] Calculated uniaxial anisotropy coefficient with increasing $d_{\mathrm{U}}$ for Fe and Ni samples. Lines are a guide to the eye.}  \label{fig:KvsDu}
\end{figure}

From the hysteresis loops, the effective uniaxial anisotropy coefficient $K_\mathrm{eff}$ was calculated using the method set out in Refs$.$ \citen{antel1999induced} \& \citen{brailsford1966physical}: the total energy density of the system, $E$ is given by
\begin{equation}
    E = -\mu_0 \Vec{H} \cdot \Vec{M} + K_\mathrm{eff}\sin^2 \gamma
\end{equation}
where $\gamma$ is the angle between the magnetization and the easy axis direction. Minimising this with respect to $\gamma$ results in
\begin{equation}\label{eqn:sat}
    H_s = \frac{2K_\mathrm{eff}}{\mu_0 M_s},
\end{equation}
when the hard axis is perpendicular to the easy axis, as assumed is the case for all samples here. In Eqn$.$ \ref{eqn:sat}, $H_s$ is the hard axis saturation field and $M_s$ is the saturation magnetization. $K_\mathrm{eff}$ was calculated assuming the active volume of magnetic material was the volume of Ni and Fe only, using the thicknesses measured from the XRR data. We also used the combined volume of the Fe and U layers for the Fe series, as illustrated in Fig$.$ \ref{fig:KvsDu}. This takes into account the possible role of an induced moment in the Fe/U samples. We note that qualitatively the two plots of $K_\mathrm{eff}$ $vs$ $d_{\mathrm{U}}$ are similar. While $K_\mathrm{eff}$ has a similar profile to those of the quantities seen in Fig$.$ \ref{fig:hcrange_fe} and \ref{fig:hcrange_ni}, there is a noticeable enhancement in the thicker samples for the Fe. When compared with Fe is is clear that the anisotropy and the resultant behaviour is weaker.

In order to understand the non-monotonic changes in the anisotropy of these two FM layers, we must examine potential variations in the microstructure of the films due to growth. It is assumed that the magnetic domain type within the FM layers does not change, therefore it is expected that the coercive field will change monotonically with roughness\cite{zhao2001effect}. In these sample sets, the roughness is approximately constant with  $d_{\mathrm U}$ for both FM series, as can be seen in Fig$.$ \ref{fig:xrr} inset for Fe, suggesting that it is not a factor in the changing anisotropy. It is also expected that any interdiffusion would be monotonic with sample thickness and therefore cannot explain the non-monotonic behavior of the anisotropy. It is also possible that there is formation of thin layers of phases such as magnetic UFe$_2$ at the Fe/U interface or UNi$_2$ for Ni/U. However, the Curie temperatures of UFe$_2$ and UNi$_2$ are 160$\,$K and 21$\,$K respectively. Therefore, if these compounds are present at the interface their magnetic properties would not contribute to the room temperature magnetic anisotropy, and their influence on the magnetic anisotropy would be monotonic with U thickness as above. As there appears to be no complete explanation for the non-monotonic anisotropy which is rooted in the material properties, we turn to electronic arguments.

Oscillations in saturation or coercive fields as a function of non-magnetic layer thickness have often been observed in heterostructures. These oscillations usually indicate a change in coupling between ferromagnetic layers. The period of this oscillatory exchange coupling of transition metals seen previously is shorter than that observed here, with Cr having the largest period of 1.8$\,$nm\cite{Parkin1991-2}. Subsequent work on heavy metal systems have shown similar oscillations periods, with Co/Pt $\sim$2$\,$nm \cite{knepper2005oscillatory}. However, the behavior observed here clearly do not fit the criteria for interlayer exchange coupling as there is no secondary FM layer to couple to. Instead we can look to quantum well states (QWSs), which while known for there importance in interlayer exchange coupling\cite{himpsel1998electronicIBM}, can also be observed in bilayer systems. These QWSs arise from confinement of electron wavefunctions at the interface, which results in the formation of standing waves. The contribution to the magnetic anisotropy from these states can come from either the ferromagnetic or non-magnetic layer\cite{przybylski, manna2013effect,wursch1997quantum}. As either layer thickness is altered, there are changes in the electronic states close to the Fermi energy of the FM, altering the magnetic anisotropy. In order for the QWSs in the non-magnetic layer to influence the magnetic anisotropy, there must be hybridization of orbitals between the layers. From XMCD measurements \cite{xmcd} it is already expected that there is strong hybridization between the Fe 3$d$ and U 5$f$ orbitals. As there is no induced moment in Ni, it may be expected that there is no hybridization and therefore we would not expect to see any non-monotonic behavior beyond the low $d_\mathrm{U}$ interfacial effects. However, it has previously been suggested that there is weak hybridization between Ni and U \cite{severin1991theoretical}, which would allow the U overlayer to influence the magnetic anisotropy of the nickel . Calculations of anisotropy energy due to QWS in Pd/Co/Pd systems find oscillations over a length scale of 20 monolayers ($\sim$ 7.5$\,$nm), with a period of 6 monolayers ($\sim$ 2.$\,$nm) \citen{cinal2001analysis}. 

As noted previously, XMCD studies on U/FM multilayers observed an induced moment in U [Ref$.$ \citen{xmcd}] when in close proximity to Fe. Wilhelm \textit{et al.} suggested that this moment is oscillatory within the U layer and its presence is a result of hybridization of Fe $3d$ and U $5f$ orbitals. Within a single U layer, the induced magnetic moment was predicted to oscillate with a period of $\sim$3$\,$nm. If we were to assume the non-monotonic behavior is indicative of oscillations, it is possible to ascribe a period similar to that of the XMCD. If we assume that the interfacial magnetization of Fe and U are locked, staying parallel or anti-parallel to one another at all angles of applied external field, then it may be expected that if the net magnetization of the U layer oscillates with thickness, then the total anisotropy of the system will concomitantly oscillate. However crucially in the previous XMCD study no moment was observed in Ni/U superlattices. Therefore if there is a connection between the induced moment in the U and the anisotropy of the Fe in the first system, it cannot be a direct causal link. A more natural solution is therefore to assume that the oscillations are driven by QWSs in the U which also influence the magnitude and sign of any induced moment, with no strong direct link between the induced moment and the FM anisotropy.

\begin{figure}[h]
 \includegraphics[scale = 1.1]{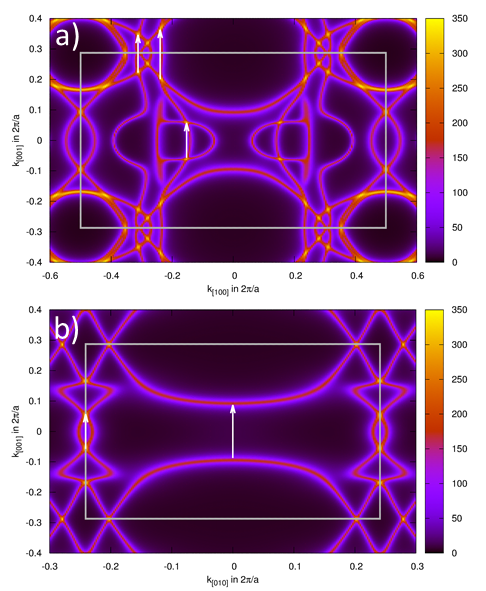}
          \caption{[Color online] Theoretical Bloch spectral functions, $S(k)$, for $\alpha$-uranium. Two-dimensional cuts are shown for lattice planes a) (010) and b) (100), respectively. The Brillouin zone boundary is shown as a grey-bordered rectangle in both images. The white arrows are indicating wave vectors connection two high density regions of the BZ and as such giving rise to possible oscillations in the region from $1.6$ to $2.5$~nm.}  \label{fig:theory}
\end{figure}

The period of oscillation, $T$, for a quantum well state in real space can be related to a wavevector of $1/T$ in reciprocal space, adapting the discussion from Ref$.$ \citen{ortega1992quantum}.  In order to quantitatively address the origin of oscillations, we have calculated the band structure of orthorhombic $\alpha$-U using density functional theory. Given the dominant (001) texture found in the samples, we look for specific features in the Bloch spectral function in the (100) and (010) planes, which include the [001] direction. The resulting Bloch spectral functions for these planes are shown in Fig$.$ \ref{fig:theory}a) and b) respectively. Notably we can find regions of relatively high spectral weight at co-ordinates $(k_x, k_y, k_z) =$ (-0.24, 0, 0.2), (-0.312, 0, 0.222), (-0.153, 0, 0.057)  and (0, -0.241, 0.057), (0, 0, 0.087) in the (010) and (100) cuts, respectively. Here the units are in terms of $2\pi/a$ with $a=2.836\;$\AA$\,$  the lattice constant of $\alpha$-U. For points close to the BZ boundary the period of oscillation is given by $\pi/(k^{[001]}_{BZ}-k_{[001]})$, with $k^{[001]}_{BZ}=\pi/c$ with $c=1.741a$, resulting in $1.6$ nm and $2.2$ nm. For the points away from the BZ boundary the oscillation is determined by $\pi / k_{[001]} $ giving periods of oscillations of $1.6$ nm and $2.5$ nm, respectively. This is in very good quantitative agreement with the roughly $2$ nm period observed in the experiment. 

This simplified analysis relies solely on the 3D band structure of the U film and as such cannot account for interfaces effects or the distinct situation in different U/FM bilayers. In order, to go one step further we can analyze the band structure of the FM materials in the dominant growth direction, Fe (011) and Ni (111). As it turns out, while for Fe both the majority and minority bands have same symmetry bands at the Fermi energy, for Ni only the minority bands cross the Fermi energy in (111) direction. This would indicate a formation of QW states in U/Ni to be more likely than in the corresponding U/Fe system. Furthermore, as indicated in the Appendix, the U films in U/Fe show different growth directions leading to a stronger averaging of any QW state periodicity.

\subsection{Out-of-plane magnetometry}
\begin{figure}[h]
 \includegraphics[scale = 0.4]{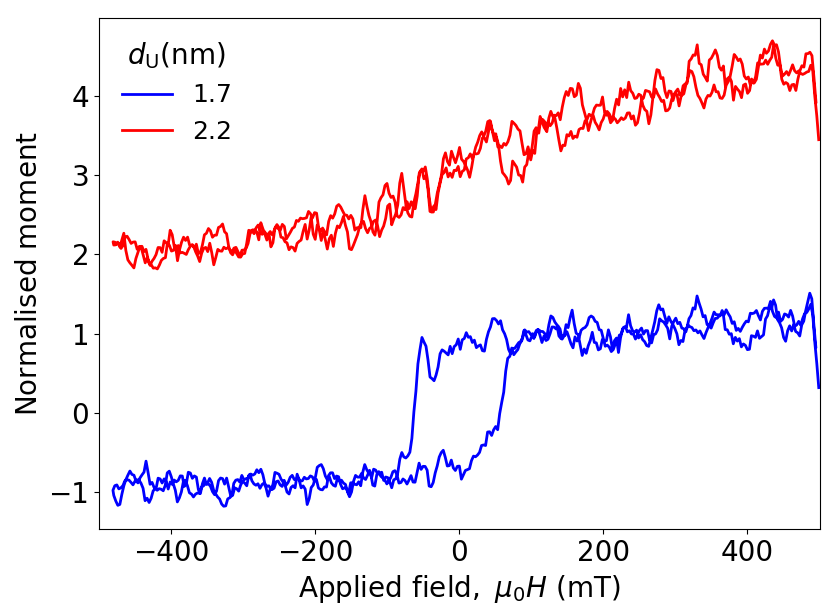}
        \caption{[Color online] Room temperature out-of-plane $M(H)$ loops for samples with $d_{\mathrm{U}} \sim$ 1.7$\,$nm and 2.2$\,$nm for the FM = Fe series. A clear hysteresis loop is observed for the sample with relatively thin U. Curves are offset vertically for clarity. }  \label{fig:oopMh}
\end{figure}

\begin{figure}[h]
 \includegraphics[scale = 0.4]{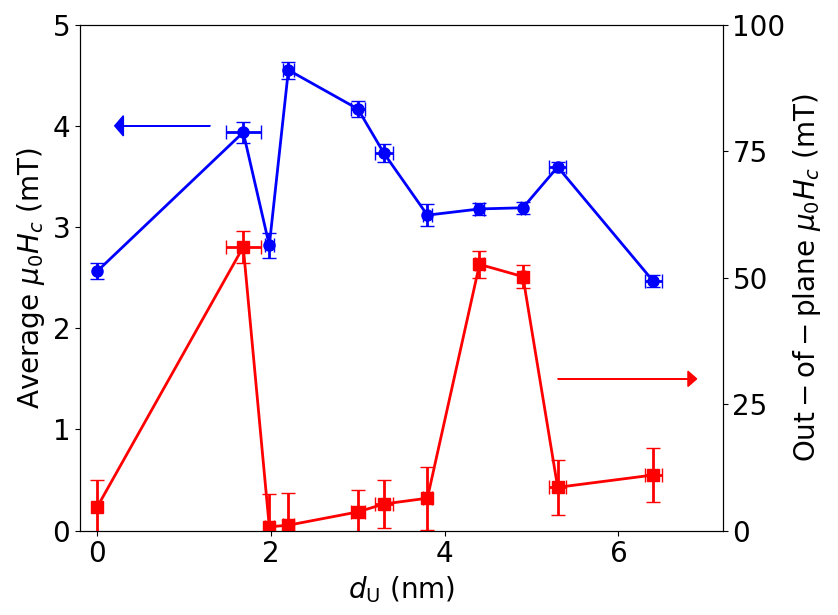}
        \caption{[Color online] Average in-plane (left) and out-of-plane (right) coercive field vs $d_{\mathrm{U}}$. }
        \label{fig:oop}
\end{figure}

Out-of-plane magnetization measurements show only hard axis behaviour for a majority of the Fe samples, and all of the Ni series. However, three Fe/U samples exhibit a clear open hysteresis loop, indicative of an easy axis response with an applied field out-of-plane: example data are shown in Fig$.$ \ref{fig:oopMh}. The out-of-plane behavior over the whole range of $d_{\mathrm U}$ is illustrated in Fig$.$ \ref{fig:oop}. It appears that the out-of-plane easy axis samples may correspond to those with lower in-plane $H_c^{\mathrm{ave}}$, however, the samples size is to small to use this as a reliable indicator.

As for the in-plane measurements, the material science arguments cannot easily explain the non-monotonic behavior of the out-of-plane anisotropy. Perpendicular magnetization is often attributed to an interfacial magnetic anisotropy, K$_s$, Ref$.$ \citen{Hashimoto1989}. Generally, samples in which perpendicular magnetic anisotropy (PMA) is observed have very thin ferromagnetic layers, on the order of < 3$\,$nm, [Refs$.$ \cite{Hashimoto1989,engle1991,johnson1996magnetic}]. To observe PMA in structures with a comparatively large FM thickness is unexpected. It may be that the presence of quantum well states also influences the out-of-plane anisotropy, though it might be expected that a continuous changes in PMA would be observed across the series rather than sudden switching.

Theoretical calculations on a number of Fe(001)/non-magnetic metal structures determined that certain metals will promote PMA when in close proximity to Fe, Ref$.$ \citen{miura2013first}. It was seen that metals with filled $d$ bands, such as Au,Ag, exhibited PMA in the Fe layer, while those with partly filled bands produced in-plane magnetization, with the exceptions on Zr and Hf. Miura \textit{et al.} suggest that PMA is observed at the Hf interface due to unoccupied majority spin $d$ states, and is enhanced by the large SOC of Hf. It is possible that this argument could be applied to uranium. However, there are two main issues in the context of the work presented in this paper. In Ref$.$ \citen{miura2013first}, the Fe film is on the order of 2.5$\,$nm (9 layers) and exhibits PMA on its own, which is not the case in this work. Secondly, if the PMA is due to the $d$ band filling, it would be expected that PMA is observed for every sample in the series.

An alternative origin of the PMA may be the interfacial Dzyaloshinskii-Moriya interaction (DMI). PMA is a generally observed in samples which exhibit interfacial DMI. Interfacial DMI is a result of large SOC of the HM layer interacting with FM spins at the interface between the two. This causes a canting of spins, pulling them out of plane. The link between interfacial DMI an induced magnetic moments has been discussed both experimentally and theoretically\cite{ryu2014chiral,yang2015anatomy,rowan2017interfacial}, with differing opinions. If the out-of-plane magnetization is indicative of interfacial DMI, then assuming the inverse relation of induced moment and DMI from the calculations of Yang \textit{et al.}, it is not unreasonable to suggest that PMA is only seen at specific thicknesses with small induced moment. However, even the largest induced moment observed in U would not be expected to overcome DMI based on the calculations by Yang \textit{et al.}. Based on the presence of PMA alone, it is not possible to draw solid conclusions on the existence of DMI within these samples. Hence, from these data alone, it is not clear whether the observed PMA is a result of a thickness dependant interfacial anisotropy in the system or interfacial DMI due to large SOC of the uranium, and further investigation is required.

\section{Conclusions}
In conclusion, both the in-plane and out-of-plane magnetic behaviour of FM/U bilayers as a function of $d_{\mathrm{U}}$ have been investigated. For both ferromagnet types the in-plane properties change in a non-monotonic manner with increasing $d_{\mathrm{U}}$. This behavior is likely linked to quantum well states formed in the uranium overlayers. Computational calculations of the Bloch spectral functions for $\alpha$-U indicate possible regions in the electronic structure which might drive oscillations which are approximately consistent with the non-monotonic data. Out-of-plane measurements revealed perpendicular magnetization for samples with thicknesses $d_{\mathrm{U}}$ =1.7, 4.4 and 5.0$\,$nm. The unexpected presence of PMA in these relatively thick films can not be easily explained and significant further study would be required to pinpoint its origin.

\section{Acknowledgements}
We thank N.-J. Steinke and G. Stenning at the ISIS Neutron and Muon Source for access to and support with the rotating anode x-ray diffractometer system. E.G. is supported by the Bath/Bristol Centre for Doctoral Training in Condensed Matter Physics, under the EPSRC (UK) Grant No. EP/L015544.  M.G. thanks the visiting professorship program of the Centre for Dynamics and Topology at Johannes Gutenberg‐University Mainz. The computational work was carried out using the computational facilities of the Advanced Computing Research Centre, University of Bristol.

\bibliography{thesis}
\bibliographystyle{apsrev}

\appendix
\section{Further x-ray data for Fe and Ni samples \label{appenda}}
\begin{figure}[h]
 \includegraphics[scale = 0.4]{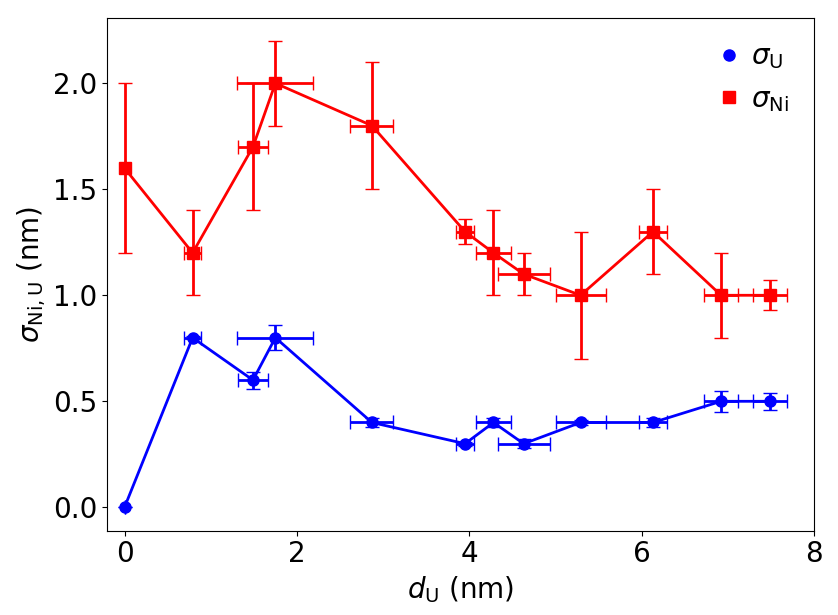}
        \caption{[Color online] Interfacial roughness for the Ni/U interface (red squares) and the U/Nb interface (blue circle) extracted from the GenX fitting to the x-ray reflectivity data, for various $d_{\mathrm{U}}$.}  \label{fig:ni_rough}
\end{figure}

X-ray reflectivity measurements were taken on all samples in the Ni-U series, in a similar way to the data shown in the main text for the Fe series. The reflectivity data could be well fitted with the GenX software, giving rise to roughness values, $\sigma_{\mathrm{U}}$ and $\sigma_{\mathrm{Ni}}$, for the U and Ni layers, respectively. These roughness values are shown versus $d_{\mathrm{U}}$ in Fig$.$ \ref{fig:ni_rough}. At relatively low $d_\mathrm{U}$ the roughness changes significantly, and there are some similarities between the form of the roughness when compared with that of the anisotropy. However, a higher thickness, where the roughness is more consistent, there is less similarity to the form of the anisotropy. This suggests that while the roughness may influence the anisotropy at lower thicknesses, it is not the mechanism which gives the anisotropy its non-monotonic form at greater thicknesses.

\begin{figure}[h]  
 \includegraphics[scale = 0.4]{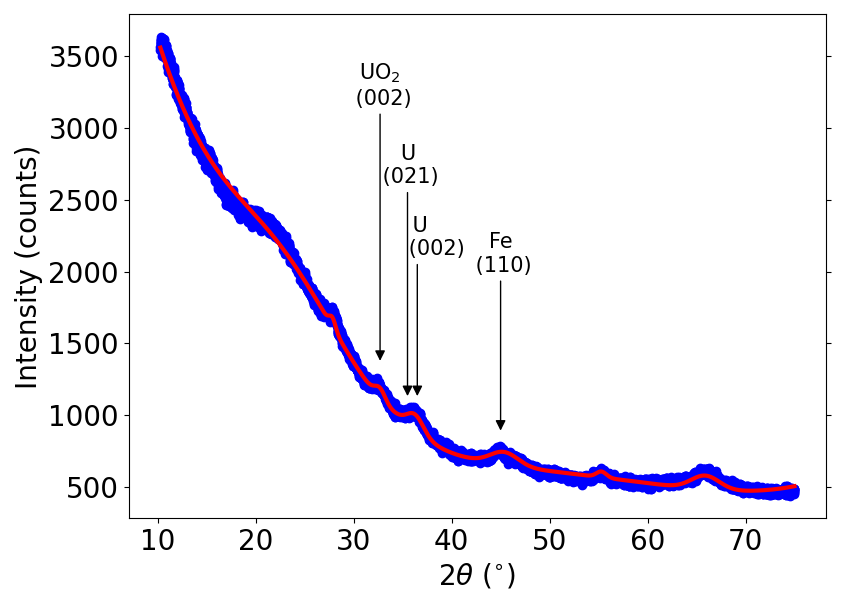}
        \caption{[Color online] GIXRD data (blue) for the Fe bilayer series with $d_{\mathrm{U}} = 3.8\,$nm. In this scan, $\omega = 1^{\circ}$. The red line is the sum of a fourth order polynomial, to represent the background (dominated by the glass substrate), and Gaussian peaks.  }  \label{fig:fe_text}
\end{figure}

\begin{figure}[h]
 \includegraphics[scale = 0.4]{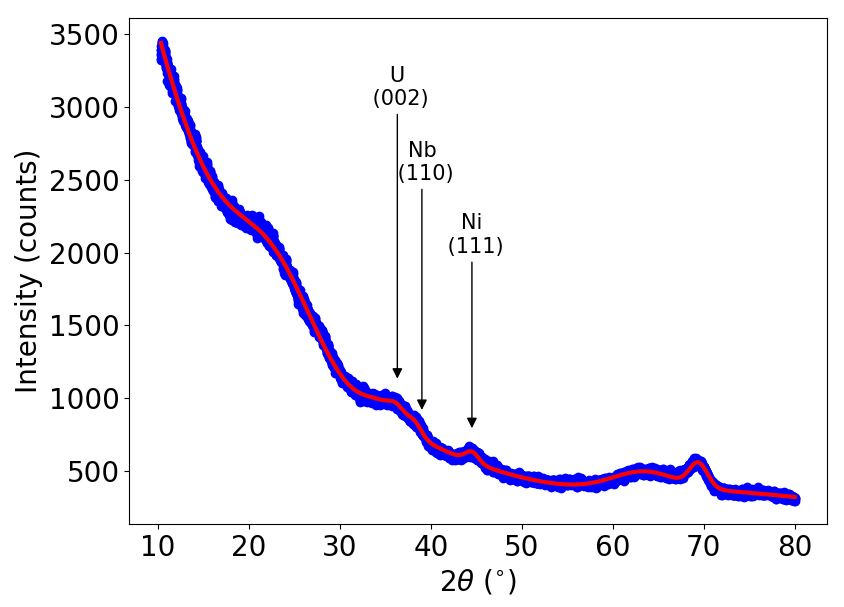}
        \caption{[Color online] GIXRD data (blue) for the Fe bilayer series with $d_{\mathrm{U}} = 6.9\, $nm. In this scan, $\omega = 1^{\circ}$. The red line is the sum of a fourth order polynomial, to represent the background (dominated by the glass substrate), and Gaussian peaks for the various layers as labeled by the arrows.  }  \label{fig:ni_text}
\end{figure}

Grazing incidence x-ray diffraction (GIXRD) scans were also carried out for representative samples in the Fe and Ni-based series. These data are shown in Figs$.$ \ref{fig:ni_text} and \ref{fig:fe_text}. In the grazing incidence geometry the incident beam angle relative to the substrate is fixed at an angle $\omega$, while the detector rotates. As $2\theta$, the angle between the incident and diffracted beam increases, the angle of the scattering vector shifts so that it always bisects the angle between the incident and detected waves. Thus, while the diffraction peaks are not necessarily parallel to the surface, the diffraction observed for small $\omega$ is still dominated by the largest component of the scattering vector, which is the normal to the thin film. Hence GIXRD provides useful information on the crystallographic texture of the samples.

The GIXRD scans were carried out over a range of $\omega$ = 0.5$^\circ$ - 10.0$^\circ$. Fits to the data are composed of a quadratic background, and a Gaussian fits to each peak. The broad uranium peak is expected to be composed of three $\alpha$-uranium peaks; (110) at 34.8$^\circ$ (021) at 35.5$^\circ$, and (002) at 36.3$^\circ$ [Ref$.$ \citen{eeles1963x}]. To determine the primary orientation of the uranium layer, these three peaks were fixed in position but allowed to vary in relative intensity and width. Fitting in this manner suggests that for Fe/U samples the U layer is a mix of [001] and [011]. For Ni, it appears that only the (002) peak is visible, suggesting orientation in only the [001] direction. 

Additional peaks were observed in both the Fe and Ni samples. In the Fe samples, a uranium oxide peak was observed. This oxidation is due to degradation of the sample over time but would not have been present in the sample at the time of the VSM measurements. In the Ni case, there is a clear Nb peak due to a thicker capping layer. The iron and nickel layers display no unexpected texture. There is no change expected in structure of the FM layers with increasing $d_\mathrm{U}$ [Refs$.$ \citen{Beesley2004,springell2008study}].

\end{document}